# AmazonNetLink: Enabling Education Access in Remote Amazonian Regions through Delay-Tolerant Networks

Andrés Fernando Barón Sandoval
*School of Computer Science*
*University of Nottingham*
Nottingham, United Kingdom
psxab21@nottingham.ac.uk

Milena Radenkovic
*School of Computer Science*
*University of Nottingham*
Nottingham, United Kingdom
milena.radenkovic@nottingham.ac.uk

*Abstract*— Access to educational materials in remote Amazonian communities is challenged by limited communication infrastructure. This paper proposes a novel delay-tolerant network (DTN) approach for content distribution and compares the Epidemic, MaxProp, and PRoPHETv2 routing protocols using the ONE simulator under dynamically changing educational file sizes. Results show that while Epidemic routing achieves higher delivery rates due to extensive message replication, it also leads to increased resource usage. MaxProp offers a balance between delivery efficiency and resource utilization by prioritizing message delivery based on predefined heuristics but struggles under high congestion and resource constraints. PRoPHETv2, with its probability-based forwarding, uses resources more efficiently but is less effective in dynamic, dense networks. This analysis highlights trade-offs between delivery performance and resource efficiency, guiding protocol selection for specific community needs. In our future work, we aim to explore adaptive buffer management and congestion-aware DTN protocols.

## I. INTRODUCTION

The biodiversity and vast territories of the Amazon present significant challenges for device connectivity and network communication systems. Establishing a comprehensive cellular network in the region entails substantial costs. The combination of high infrastructure expenses and low population density across large areas diminishes the economic appeal for telecommunications providers. This financial reality limits their incentive to invest in expanding services to these remote areas, where potential returns may not justify the expenditure [1].

Limited cellular connectivity has serious implications for the indigenous and local communities residing in the Amazon. Reliable access to mobile networks is essential for daily life as it facilitates communication, information dissemination, and the delivery of critical services such as healthcare and education. Without adequate connectivity, these communities experience significant isolation, which hampers their ability to engage with the wider world, access educational resources, or receive modern medical consultations and support. This digital exclusion exacerbates social and economic inequalities, leaving these populations disconnected from advancements that could enhance their quality of life and support community development. According to a report by the Fundación Empresarios por la Educación, a prominent Colombian organization aimed at positively impacting the educational sector in various regions [2], 86% of the indigenous population in Colombia faces significant challenges in providing education that aligns with the culture, context, and worldview of its indigenous population. Notably, 30% of this population lacks access to formal education, contributing to an illiteracy rate of 32% among indigenous communities. This research explores communication strategies based on delay-tolerant networks (DTNs) using Epidemic, MaxProp, and PRoPHETv2 protocols to identify the most effective approach for optimizing access to educational resources in these regions analyzing delivery probability, latency, and resource efficiency. The objective is to mitigate educational disparities and foster greater inclusivity through the strategic application of technology.

## II. RELATED WORK

Delay-Tolerant Networks (DTNs) differ fundamentally from traditional cellular networks as they do not rely on a predefined end-to-end path for packet delivery. Instead, they use a 'store-carry-forward' mechanism, where nodes temporarily store packets in their buffer, carry them while in motion, and forward them when connectivity opportunities arise. To optimize this process, various communication protocols have been developed to enhance packet delivery performance. We are particularly interested in benchmarking core protocols that show high potential for addressing this problem

MaxProp enhances message delivery in Delay-Tolerant Networks (DTNs) by ranking packets based on their likelihood of reaching their destination. This prioritization enables efficient message transmission and deletion, optimizing resource use. With acknowledgment mechanisms and proactive buffer management, MaxProp improves reliability and network efficiency [3].

PRoPHETv2 optimizes routing by leveraging predictable movement patterns. Nodes exchange and update delivery probabilities during encounters, refining routing decisions over time. By prioritizing nodes with higher transmission potential, PRoPHETv2 excels in dynamic networks with intermittent connectivity and high delays [4].

Epidemic routing ensures message propagation through uncontrolled flooding—nodes exchange all unseen messages upon contact, creating multiple copies across the network. While maximizing delivery probability, this approach consumes significant bandwidth, storage, and energy, making it ideal for high-reliability scenarios like disaster response and remote communications.

*A. Edge Storage in DTN Routing*

In Delay-Tolerant Networks (DTNs), where communication is frequently disrupted, effective message delivery depends on efficiently managing limited buffer space. Unlike traditional networks, DTNs rely on intermediate nodes to store messages for extended periods, making buffer capacity essential. [6]. By developing buffer-aware routing protocols that dynamically adjust forwarding decisions based on real-time buffer conditions, DTNs could prioritize urgent messages, prevent data loss, and reduce congestion, ultimately improving network reliability. This approach allows DTNs to adapt to resource-constrained environments—such as rural and disaster-stricken areas—by tailoring message handling to the specific capacities of each node, thereby optimizing delivery rates and ensuring that essential information reaches its destination [6].

*B. Contact Volume in DTNs*

The communication problem in the Amazon has been explored in previous research. An article presented at ExtremeCom discusses the Amazon Regatão project [7], which aims to provide Internet access to riparian communities along the Amazon River using Delay-Tolerant Network (DTN) schemes. This project developed a technique to predict the number of data packets (bundles) that can be transferred during a network contact, optimizing transmission through proactive fragmentation strategies and queue management. This research is particularly valuable as it improves the efficiency of DTN networks in contexts where connectivity is intermittent, and communication resources are limited. Furthermore, it provides an experimental framework based on real-world data that can be adapted or extended to address specific needs in remote environments, such as developing solutions for the transmission of essential information to isolated communities. By building on the methodologies and findings of the Amazon Regatão project, this work aims to evaluate and refine protocols to ensure reliable and efficient delivery of educational materials, tailored to the unique requirements of these regions.

### III. ENHANCING ACCESS TO EDUCATIONAL RESOURCES VIA PEER-TO-PEER FILE SHARING IN AMAZONIAN COMMUNITIES

*A. Motivation and Background of the Geospatial Scenario*

Currently, around 385 Indigenous groups reside across approximately 2.4 million square kilometers (about 930,000 square miles) of the Amazon, maintaining diverse cultural traditions and ways of life [9]. However, access to education remains a significant challenge, as about 42% of children and adolescents aged 6 to 16 struggle to fully participate in remote learning due to limited electricity, internet connectivity, and technological infrastructure. This digital divide has led to high dropout rates, forcing many students to abandon their studies for the school year, further exacerbating educational disparities and limiting future opportunities for Indigenous youth [10].

Indigenous communities around the municipality of Leticia in the Amazonas region form the core focus of the investigation. Most of these communities are located near the banks of the Amazon River, close to the Amacayacu National Natural Park, one of the most important protected areas in the region. Some communities, such as the inhabitants of Mocagua, belong to the Tikuna ethnic group, one of the largest and oldest Indigenous communities in the Amazon, recognized for their rich cultural heritage, native language, and ancestral traditions [11].

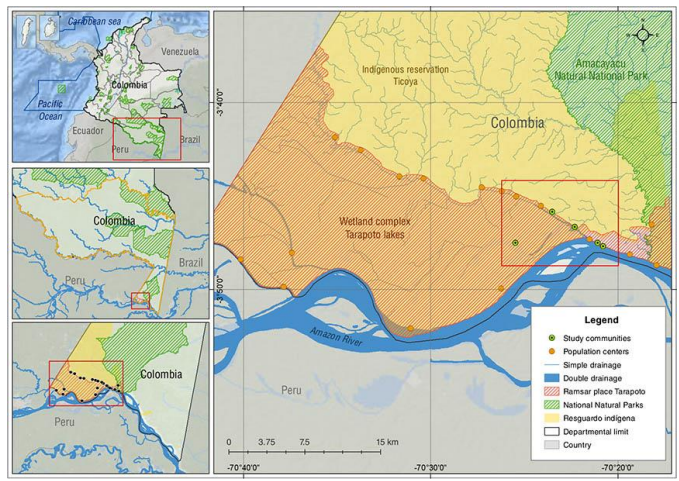

Fig 1. Map of several indigenous communities in the Amazon jungle, located in the southern region of Colombia [12].

After identifying the geographic location of the surrounding communities, it became evident that developing a communication strategy independent of cellular networks was essential due to the area's isolation within the dense jungle and its limited road access.

These communities rely heavily on fishing in the Amazon Basin for both subsistence and economic activities, selling their catch in local markets, while ecotourism also contributes to their livelihood by attracting visitors interested in the region's biodiversity and indigenous culture. Within the towns, bicycles and walking are the primary modes of transportation due to the lack of vehicular infrastructure, preserving the natural landscape. For longer distances, wooden canoes, motorized boats, and traditional paddle canoes serve as essential means of transport for trade, education, and healthcare access, reinforcing the Amazon River's central role as a natural highway that sustains daily life, economic exchanges, and community connectivity.

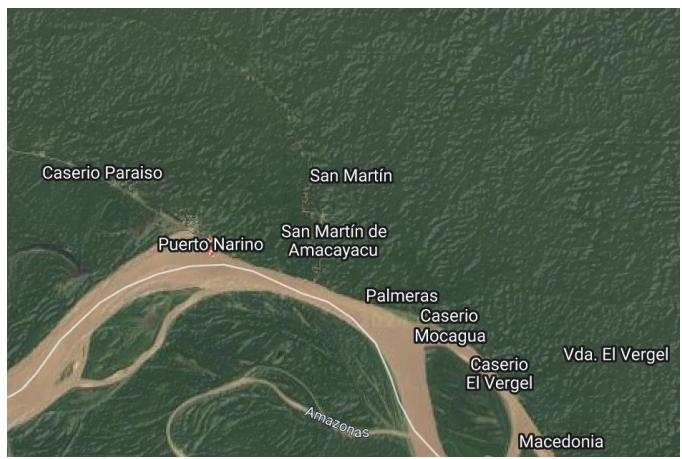

Fig. 2. Satellite view showing Puerto Nariño and nearby towns nestled in the dense jungle of Leticia, Colombia.

Based on detailed analysis of data and geographic coordinates, a map was created to model the primary water basins where fishermen and touristic transportation systems converge, as illustrated in Figures 3 and 4.

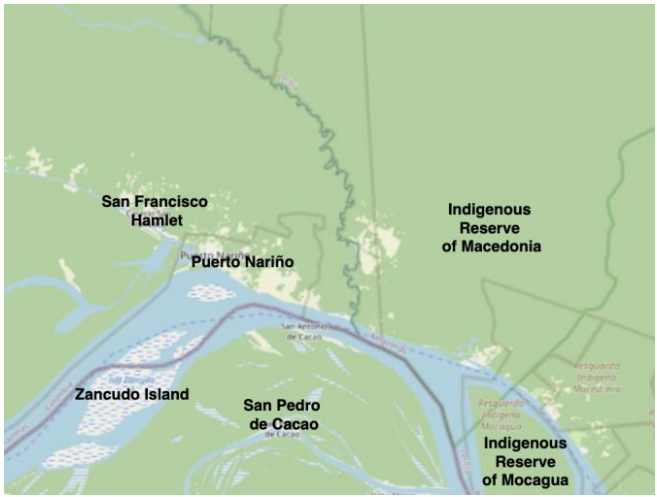

Fig 3. Map of key Indigenous Reserves and Settlements in the Amazon Region.

In Figure 4, the community of Puerto Nariño is not visible due to the scale of the image, which limits the level of detail for smaller locations. Puerto Nariño is a small yet significant municipality in the Amazonas Department of southern Colombia. Known as the "Ecological Capital of Colombia" due to its commitment to sustainability and environmental preservation, the town prohibits motorized vehicles, promoting a harmonious relationship between the community and its natural surroundings. Situated along the Amazon River, Puerto Nariño serves as a vital hub for numerous indigenous communities, providing access to essential resources, healthcare, education, and cultural exchanges.

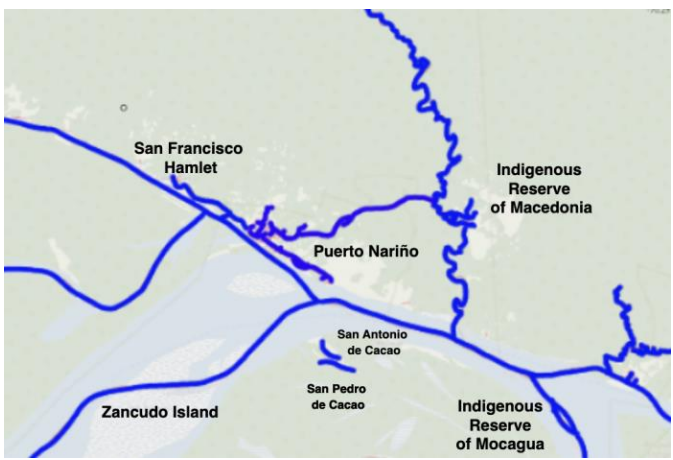

Fig 4. Map of Key Indigenous Reserves and Settlements in the Amazon Region Near Puerto Nariño, Including Principal Roads and Waterways in the Area [13].

Puerto Nariño's strategic riverine location makes it a key point for transportation, trade, and communication, linking remote settlements with larger urban centers while preserving the region's ecological integrity.

For a clearer view of Puerto Nariño and its surroundings, a more detailed representation is provided in Figure 5.

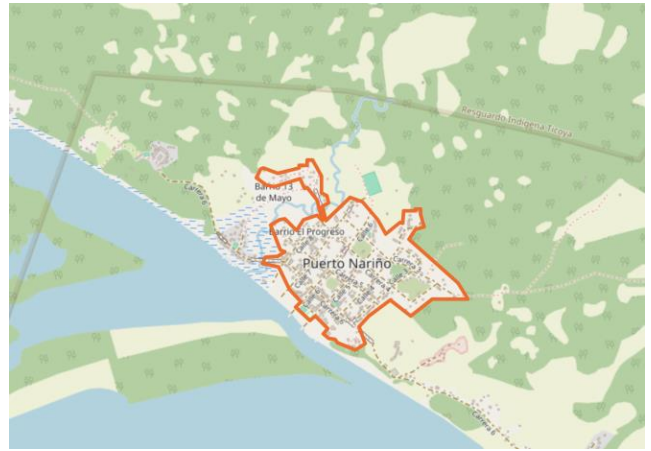

Fig. 5. Map of Puerto Nariño, a central hub for indigenous convergence and neighboring communities. [13]

By identifying the maritime and terrestrial communication routes, the map presented in Figure 6 was obtained.

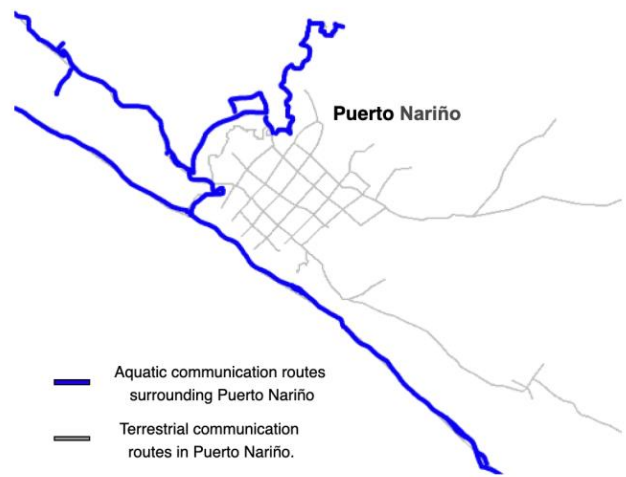

Fig. 6. Map of Puerto Nariño, showing principal terrestrial and aquatic roads, suitable for communication in the area [13].

*B. Design and build of AmazonNetLink*

Due to the limited transportation infrastructure in the Leticia and Puerto Nariño region, the experiment incorporates the traditional means of communication relied upon by the local population, namely bicycles for land travel and motorboats for navigation along waterways. These modes of transportation are deeply rooted in the daily lives of the residents, given the lack of vehicular traffic and the absence of roads connecting many areas. Bicycles are the primary means of mobility within small cities like Puerto Nariño, while motorboats play a crucial role in connecting communities along the Amazon River and its tributaries, serving as lifelines for goods, services, and communication. This decision ensures the experiment reflects the region's unique mobility patterns, capturing the interplay between terrestrial and aquatic routes and enabling an accurate representation of how educational resources could realistically be transmitted across this geographically complex and infrastructure-limited environment.

Based on available literature and community data, it is estimated that an average of approximately 70 fishermen operate daily, engaging in fishing activities for about eight hours, which aligns with the typical workday in Colombia [14]. No specific data was available regarding the number of bicycles in the main city (Puerto Nariño) and the surrounding indigenous communities. Therefore, an average of 50 bicycles was assumed for the analysis.

To evaluate the feasibility of transmitting educational resources under these conditions, each protocol was executed individually in The ONE simulator. The ONE (Opportunistic Network Environment) simulator is a research tool for Delay-Tolerant Networks (DTNs), supporting mobility models and routing protocols such as Epidemic, PRoPHET, and MaxProp. It enables realistic simulations by importing real-world mobility traces and provides visualizations of node behavior and performance metrics.

The simulation lasted 43,200 seconds (12 hours) and modeled data transfers aligned with the region's mobility patterns. It used average file sizes ranging from 0 MB to 3 MB, suitable for sharing instructional videos, short tutorials, infographics, presentations, PDF guides, and visual aids such as diagrams and photos. These simulations aim to determine how effectively the protocols operate within the constraints imposed by the local geography and transportation methods.

After processing the collected data, the following scenario was obtained, highlighting indigenous shelters and the city of Puerto Nariño

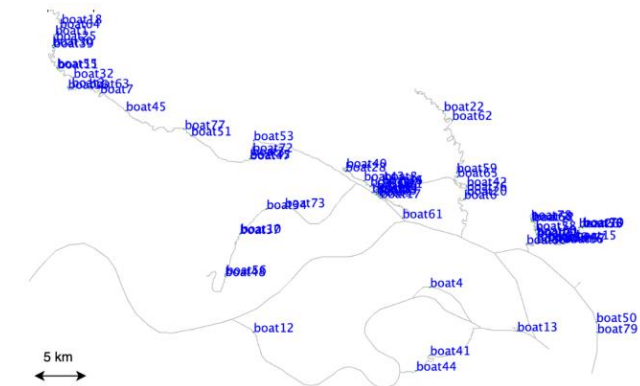

Fig. 7. Movement Nodes in a Pseudo-Realistic Experiment Given the Experimental Constraints for Economic Activities Near Puerto Nariño

As shown in Figure 8, some clusters of nodes are clearly visible, forming dense connectivity regions within the network. This clustering is advantageous for communication, as a higher concentration of nearby transfer nodes increases the likelihood of successful data delivery, minimizes delays, and enhances message redundancy. Additionally, these clusters serve as natural relay points, improving routing efficiency by reducing the number of hops needed for transmission. In scenarios with intermittent connectivity, such as those common in Delay-Tolerant Networks (DTNs), these high-density areas play a critical role in mitigating disruptions, ensuring that messages have multiple potential forwarding paths, thereby improving overall network resilience and performance.

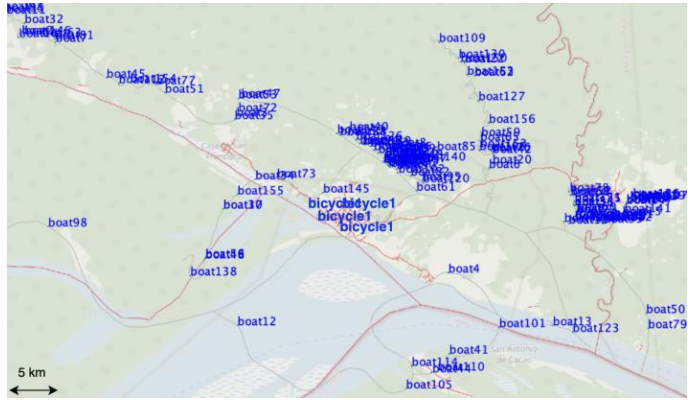

Fig. 8. Movement nodes in a pseudo-realistic experiment given the experimental constraints for economic activities near Puerto Nariño with background map of the area.

Initially, the transmission speed was intended to be a constant value. However, to better represent the conditions of the Amazon River, the ONE simulator was modified to handle data transmission at a variable rate. This change was implemented in the SimpleBroadcastInterface, a class responsible for packet transmission during the simulation. A variable, dynamicTransmitSpeed, was defined and dynamically adjusted throughout the simulation. It was initially set to 500 KBps (kilobytes per second).

```
private double dynamicTransmitSpeed = 500 * 1000;
```

Fig. 9. Dynamic Definition of the Data Transmission Variable

The method responsible for message transmission canTransmit() is updated to utilize dynamicTransmitSpeed, introducing logic to modify the transmission speed dynamically based on predefined conditions, such as simulation time or node interactions.

```
@Override
public boolean canTransmit() {
    updateDynamicSpeed();
    return (currentTime - lastTransmitTime) >= (1 /
        dynamicTransmitSpeed);
}

private void updateDynamicSpeed() {
    if (SimClock.getTime() > 10000) {
        dynamicTransmitSpeed = 1 * 1000 * 1000;
    }
}
```

Fig. 10. Definition for the updateDynamicSpeed function

In the canTransmit() method, a check is performed to determine if sufficient time has elapsed since the last transmission. This is dynamically managed by invoking the updateDynamicSpeed() method. The updateDynamicSpeed() implementation increases the transmission speed to 1 Mbps once the simulation clock exceeds 10,000 seconds. This dynamic adjustment allows the simulation to more accurately represent real-world scenarios where transmission speeds can vary due to environmental changes or system demands, thereby enhancing the realism and adaptability of the network simulation.

Table 1 summarizes the key configuration settings used in the AmazonNetLink experiment. The interface settings establish the primary parameters for communication, including a variable transmission speed of 500 kbps, a transmission range of 3 meters, a buffer size of 8 MB, and a latency of 200 milliseconds.

Group 1 (Bicycles) consists of 50 hosts, each moving at a constant speed of 1 m/s, reflecting the slower pace of terrestrial transport within small towns like Puerto Nariño.

Group 2 (Motorboats) represents faster aquatic transportation and includes 70 hosts moving at a significantly higher speed of 15 m/s. This distinction models the increased transit efficiency of motorboats along river networks, which are vital for connecting remote communities.

Bluetooth was chosen as the communication interface for this experiment due to its low power consumption, simplicity, and widespread adoption in resource-constrained devices. These characteristics make it an ideal starting point for simulating communication in scenarios where infrastructure is limited, and energy efficiency is crucial.

| Category | Settings |
|---|---|
| Interface Settings (Bluetooth-Based Communication) | |
| Transmission Speed | 500 kbps (Variable) |
| Transmission Range | 3 meters |
| Buffer Size | 8 MB |
| Latency | 200 ms |
| Group 1: Bicycles | |
| Movement Model | ShortestPathMapBasedMovement |
| Number of Hosts | 50 |
| Speed | 1 m/s |
| Group 2: Motorboats | |
| Movement Model | ShortestPathMapBasedMovement |
| Number of Hosts | 70 |
| Speed | 15 m/s |

Table 1: Experiment configuration settings common for the three protocols.

Unlike other protocols, PRoPHETv2 requires additional parameters for its unique routing system, which can be expressed mathematically through the delivery probability equation.

$$P_{A,B} = P_{A,B} + (1 - P_{A,B}) \cdot pInit$$

In this equation:
- $P_{A,B}$ corresponds to the probability between nodes A and B.
- pInit is the initial probability assigned when two nodes encounter each other.

As time progresses, the probability of successful message delivery decreases, reflecting the declining significance of the information.

$$P_{A,B} = P_{A,B} \cdot \beta^{\Delta t}$$

In this equation:
- $\beta$ represents the decay factor based on time.
- $\Delta t$ denotes the duration since the most recent encounter.

By incorporating intermediate nodes into the routing calculation, the transitivity rule optimizes the likelihood of successful message delivery:

$$P_{A,C} = P_{A,C} \cdot (1 - P_{A,C}) \cdot P_{A,C} \cdot P_{B,C} \cdot \gamma$$

In this equation:
- $\gamma$ represents the influence of an intermediate node B in facilitating the connection between A and C.
- $P_{A,C}$ corresponds to the probability between nodes A and C.

Given the previous relationship to dynamically adjust delivery probabilities, the values assigned for the simulation are as follows:

| Parameter | Value |
|---|---|
| ProphetRouter.pInit | 0.75 |
| ProphetRouter.beta | 0.25 |
| ProphetRouter.gamma | 0.98 |
| ProphetRouter.secondsInTimeUnit | 30 |

Table 2: PRoPHETv2 Router Configuration Parameters

*C. Analysis of Geo-temporal Spatial Dynamics*

Observing the results in Fig. 11, it can be noted that Epidemic exhibits a higher delivery probability compared to MaxProp and PRoPHETv2 for file sizes smaller than approximately 1.7 MB. However, for larger file sizes, Epidemic's performance begins to decline rapidly, and for files of 2 MB or more, both MaxProp and PRoPHETv2 demonstrate significantly higher delivery probabilities than Epidemic.

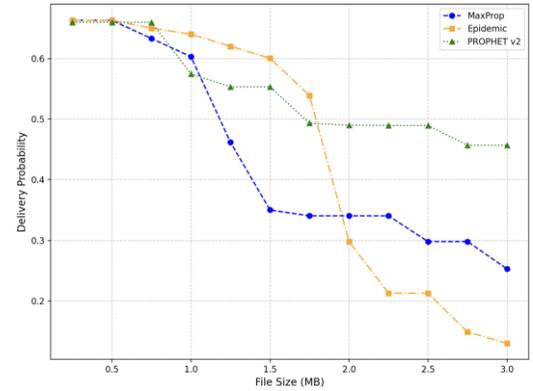

Fig. 11. Delivery Probability vs File Size (MB) for the three protocols analyzed

This rapid decline can be attributed to the flooding-based mechanism of Epidemic, which indiscriminately replicates messages to all encountered nodes in an attempt to maximize delivery probability. While this strategy works well for small file sizes due to the lower resource requirements, it becomes inefficient as file sizes increase. Larger files consume significantly more buffer space and bandwidth, quickly saturating the nodes' available resources. As file sizes increase, MaxProp's delivery probability decreases, highlighting its difficulty in handling larger files compared to PRoPHETv2. This disparity becomes particularly evident beyond 2 MB, where PRoPHETv2 consistently outperforms MaxProp. PRoPHETv2's probabilistic routing approach enables more efficient routing decisions under high resource demand. In contrast, MaxProp's prioritization strategy is less effective for larger file sizes.

Figure 12 compares Average Latency and File Size for the three protocols. As file size increases, average latency grows slightly for all protocols due to greater resource demands. MaxProp exhibits the highest latency, likely due to its cost-based prioritization introducing delays. Epidemic shows lower latency than MaxProp but still lags behind PRoPHETv2, as its flooding-based nature leads to congestion and buffer contention. PRoPHETv2 achieves the lowest latency across all file sizes, minimizing unnecessary transmissions and delays [15].

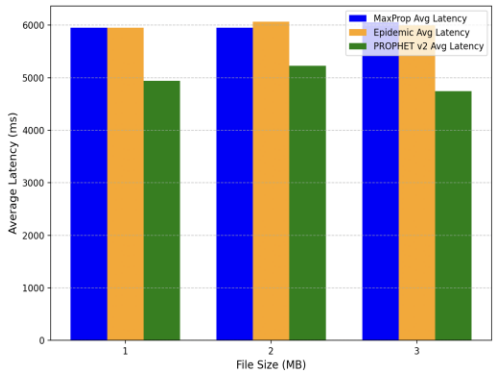

Fig. 12. Average Latency (ms) vs File Size (MB) for the three protocols analyzed

Figure 13. shows the Overhead Ratio as a function of File Size (MB) for three DTN routing protocols. The overhead ratio measures the additional transmissions required to deliver messages compared to the ideal number, providing insight into the resource efficiency of each protocol. For Epidemic, while achieving high delivery probabilities, incurs excessive overhead, limiting its practicality in large-scale or resource-constrained networks. MaxProp shows a consistently lower overhead ratio than Epidemic but remains significantly higher than PRoPHETv2. PRoPHETv2 achieves the lowest overhead ratio throughout, indicating better resource utilization.

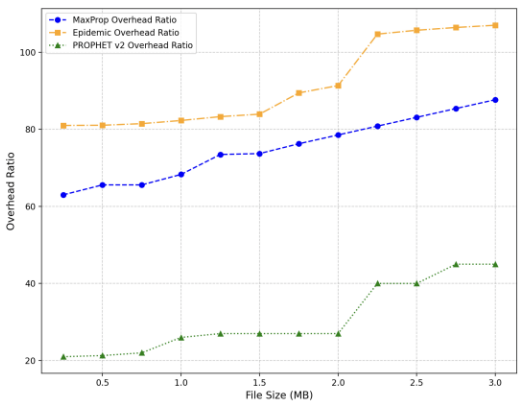

Fig. 13. Overhead Ratio vs File Size (MB) for the three protocols analyzed

Analyzing the results presented in Fig. 14, Fig. 15, and Fig. 16, it can be concluded that there is a negative relationship between overhead ratio and delivery probability. Specifically, the graphs show that as the overhead ratio increases, the delivery probability decreases.

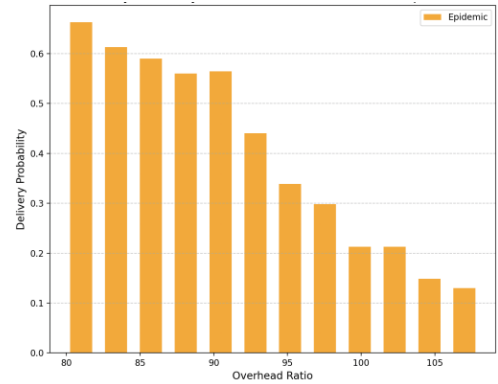

Fig. 14. Delivery Probability vs Overhead Ratio (Epidemic)

Furthermore, Fig. 15 shows the MaxProp model appears to be more sensitive to changes in overhead ratio compared to the Epidemic model. The MaxProp model exhibits a steeper decline in delivery probability as the overhead ratio increases, indicating it is more affected by higher overhead ratios.

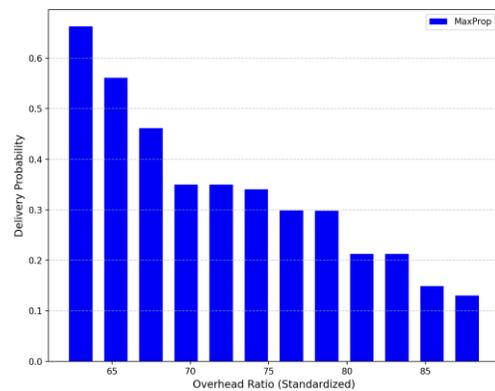

Fig. 15. Delivery Probability vs Overhead Ratio (MaxProp)

PRoPHETv2 model seems to exhibit a more moderate sensitivity to overhead ratio changes, suggesting it could be a suitable choice in scenarios where maintaining acceptable delivery probability is important even as network overhead increases. Its performance characteristics appear distinct from the steeper decline seen in the Epidemic and MaxProp models.

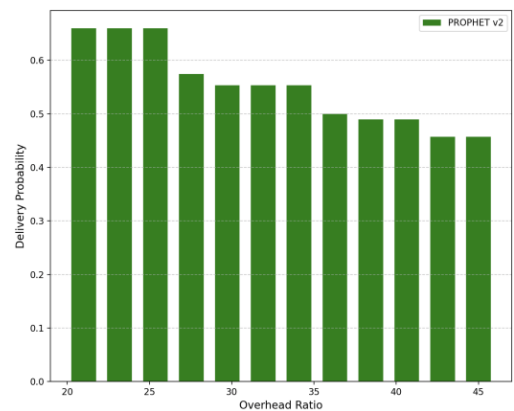

Fig. 16. Delivery Probability vs Overhead Ratio (PRoPHETv2)

## IV. DISCUSSION OF RESULTS

The analysis of the routing protocols—MaxProp, Epidemic, and PROPHET v2—under varying file sizes in a Delay-Tolerant Network (DTN) highlights notable differences in performance, particularly regarding overhead ratio and delivery probability. Epidemic consistently exhibits the highest overhead ratio, nearing 100%, as its transmission method leads to significant redundancy and network congestion, requiring substantial resources. While it ensures high delivery probabilities for small file sizes, its effectiveness diminishes drastically as file size increases, reflecting poor scalability. MaxProp achieves a moderate overhead ratio and outperforms Epidemic for larger file sizes but shows a steep decline in delivery probability as file sizes grow, indicating limited efficiency in handling larger transfers. In contrast, PRoPHETv2 demonstrates superior performance across both metrics, maintaining the lowest overhead ratio and the highest delivery probability, particularly for larger file sizes.

The results from the experiment highlight how buffer space constraints significantly influence latency in Delay-Tolerant Networks (DTNs). As file sizes increase, latency rises across all protocols, reflecting the challenge of managing larger data packets within limited buffer capacities. Epidemic demonstrates the highest latency, likely due to buffer saturation caused by its message duplication strategy, which overwhelms available resources. MaxProp, while slightly better, still shows increased latency as file sizes grow, indicating that its prioritization approach does not fully alleviate buffer constraints. In contrast, PROPHET v2 achieves consistently lower latency, suggesting more efficient buffer utilization and better handling of larger files. These findings emphasize the importance of optimizing buffer management to minimize delays and maintain reliable performance, particularly in scenarios with constrained resources and increasing data demands.

The efficiency of PRoPHET v2 is especially advantageous for applications like educational content distribution, where reliable delivery and minimal delays are essential for usability. Its ability to maintain low overhead and high delivery probability ensures that PROPHET v2 can scale effectively in networks with limited bandwidth or high variability, making it a robust solution for environments requiring consistent performance, such as remote or resource-constrained educational deployments.

Overall, among the protocols analyzed, PRoPHET v2 emerges as the most promising for transmitting files of various sizes in remote and challenging environments with little to no cellular network connectivity. As evidenced in these experiments, the limited communication resources of devices, such as memory capacity, make it essential to justify the choice of protocols for each specific use case. In this context, the use of Delay-Tolerant Networks (DTNs) could be a pivotal strategy to drive advancements in these regions, as they enable efficient data transmission in environments with intermittent connectivity. DTNs could facilitate the implementation of technological solutions such as educational platforms, telemedicine services, and community communication systems that function reliably even under adverse conditions. Communities in the Amazon, despite their late integration into the modern world, require an adequate transition, which must be supported by governments through investments in technological infrastructure, training programs, and collaboration with scientific initiatives. This would ensure fair access to essential goods such as education and communication, fostering equitable and sustainable development in these regions.

## V. FUTURE WORK

AmazonNetLink represents the first stage of research focused on the transmission of educational resources in the Amazonian region. To establish a baseline, benchmarks were conducted using three well-known protocols commonly employed in both industry and academia for Delay-Tolerant Network (DTN) communications. Based on these results, PRoPHETv2 was identified as the most suitable protocol for the scenario under consideration. However, further enhancements to this protocol are possible, such as those proposed by Ho-Jong Lee and Jae-Choong Nam [16]. Their research incorporates contact duration and message transfer time into the calculation of delivery probabilities, addressing key limitations of existing routing strategies. While PRoPHETv2 mitigates the issue of inflated delivery probabilities caused by frequent short-term encounters through the use of inter-meeting time, it does not consider the duration of node contact, a critical factor in DTNs where contact durations are often brief due to mobility and limited communication ranges [16]. By enhancing PRoPHETv2 with these considerations, delivery probability calculations become more accurate, prioritizing messages based on realistic delivery opportunities [16].

In addition to PRoPHETv2, other advanced protocols are specifically optimized for congestion awareness and replication, such as Café and CafRep[17]. Café is an adaptive forwarding protocol designed to address congestion in DTNs by diverting traffic away from centrality-driven paths prone to congestion. It uses heuristics to identify and route traffic through less congested areas. However, its primary limitation lies in its inability to reduce the overall traffic load in the network, as it focuses on redirecting traffic rather than decreasing it. This approach can be suboptimal when alternative, uncongested paths are unavailable [17].

Building on the limitations of Café, CafRep offers a unified congestion control framework for DTNs by integrating adaptive forwarding and adaptive replication management techniques. This framework not only redistributes traffic to less congested parts of the network but also actively reduces the total network traffic by dynamically adjusting forwarding and replication behaviors. CafRep employs heuristics based on node centrality, contact analysis, resource availability, and network congestion to optimize message dissemination effectively [17].

CognitiveCache is another interesting protocol that is worth mentioning for future benchmarking and discussion in the protocol selection process. This protocol employs a multi-agent deep reinforcement learning (DRL) framework to optimize content caching at network edges, enabling adaptive and collaborative decision-making among distributed nodes such as access points, mobile devices, and femtocells [18]. Its ability to dynamically adjust caching strategies based on spatial-temporal locality and user content request patterns makes it highly relevant for scenarios involving the transmission of educational resources. By improving cache hit ratios, reducing latency, and lowering transmission costs, CognitiveCache could ensure more efficient and reliable delivery of educational materials in resource-constrained and connectivity-limited environments, such as remote Amazonian communities.

To further explore its applicability, AmazonNetLink is set to be integrated into MODiToNeS [19], a low-cost, lightweight, open-source testbed designed for rapid prototyping and testing of real-time, multi-hop communication protocols in dynamic environments. This deployment builds on previous work in Valencia [20], where MODiToNeS was successfully utilized in vehicular and drone-based edge clouds across urban and agricultural settings in Spain and the UK, supporting disconnection-tolerant vehicular networks. This integration will enable the evaluation of topology and mobility control, while also facilitating predictive analysis of connectivity patterns, transmission performance, and network adaptability.

Additionally, RasPiPCloud, a lightweight mobile personal cloud framework, will be incorporated into MODiToNeS. RasPiPCloud supports real-time data collection, storage, and processing in resource-constrained environments, providing a privacy-aware, decentralized infrastructure for opportunistic networking [20]. By leveraging this platform, AmazonNetLink's adaptability can be enhanced, ensuring efficient edge processing and secure content distribution in areas where conventional cloud services are unreliable.

Inspired by the emerging research on energy aware opportunistic charging and energy distribution and sustainable vehicle edge and fog networks[21], in our future work we will aim to ensure the sustainability and economic efficiency of our proposal. Such analysis will help identify optimal resource allocation strategies, reduce operational costs, and maximize the impact on remote communities. Assessing the scalability of the project will contribute to its potential replication in other environments with similar conditions, ensuring a comprehensive approach that integrates technical, economic, and social considerations.